\documentclass{PoS}
\usepackage{lineno}
\usepackage{float}
\usepackage{url}
\title{Calibration and performance of the CMS pixel detector in LHC Run 2}

\ShortTitle{Calibration and performance of the CMS pixel detector in LHC Run 2}

\author{\speaker{Tamas Almos Vami} for the CMS Collaboration\\
        Wigner Research Centre for Physics\\
        E-mail: \email{tamas.almos.vami@cern.ch}}

\abstract{The Compact Muon Solenoid (CMS) is one of two general-purpose detectors that reconstruct the products of high energy particle interactions at the Large Hadron Collider at CERN. The silicon pixel detector is the innermost component of the CMS tracking system. It determines the trajectories of charged particles originating from the interaction region with high resolution enabling precise momentum and impact parameter measurements in the tracker. It is designed to operate in the high particle density environment of the LHC. The calibration of the pixel detector plays an important role in its performance. The calibration constants follow the physical changes in the sensors that are mostly induced by irradiation. These constants are regularly updated, maintained in a calibration database and used for the event reconstruction. We will present details on the offline calibration procedures and their effects on detector performance during the Run 2 period of LHC.}

\FullConference{7th Conference of Large Hadron Collider Physics\\
		(LHCP2019)\\
		20 - 25 May, 2019\\
		Puebla, Mexico}

\begin{document}

\section{Introduction}
The silicon pixel detector is the innermost component of the CMS tracking system. It determines the trajectories of charged particles originating from the interaction region with high resolution enabling precise momentum and impact parameter measurements in the tracker. It is designed to operate in the high particle density environment of the LHC.

\begin{figure}[h!]
    \centering
    \includegraphics[width=\textwidth]{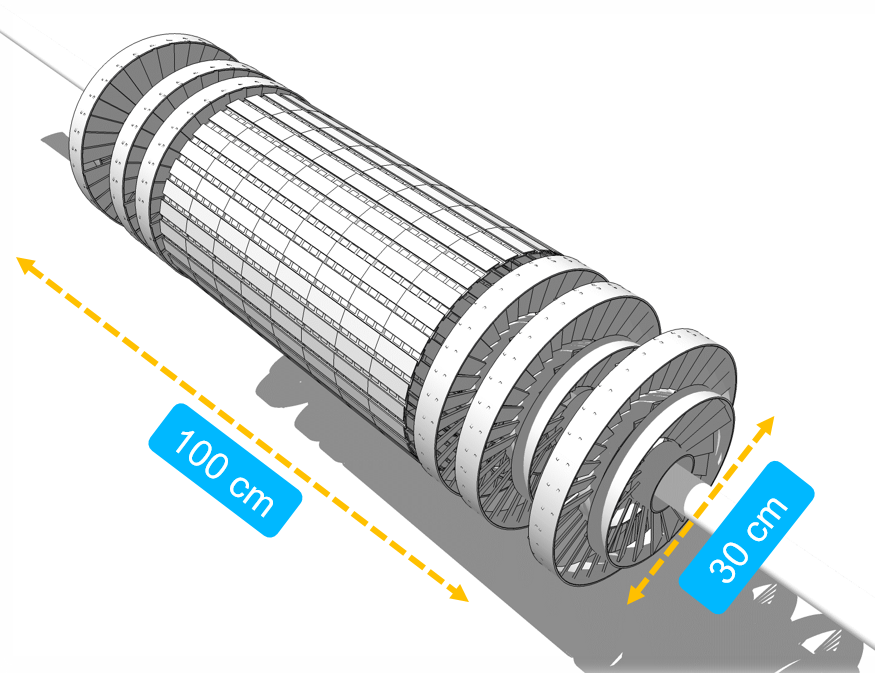}
    \caption{CMS Phase-1 Pixel Detector.}
    \label{fig:Fig1}
\end{figure}

In 2017 the original detector was replaced with a newly constructed detector (called the Phase-1 upgrade), shown in Fig.\,\ref{fig:Fig1}. The Phase-1 pixel detector has four concentric barrel layers and three disks per side containing 124 million pixels with a size of 100$\times$150\,$\mu$m$^2$. The sensor thickness amounts to 285\,$\mu$m.

\section{Calibrations and Local Reconstruction}
The presence of a pixel detector hit is determined from pixels with deposited charge above a certain threshold. Then adjacent pixels are combined into clusters. The cluster charge and position is then used to determine the hits.

The charge carriers inside the silicon bulk are deflected by the Lorentz force due to the 3.8 T magnetic field of the CMS detector. Such deflection is characterized by the Lorentz Angle ($\theta_{LA}$).

\begin{figure}[h!]
    \centering
    \includegraphics[width=.6\textwidth]{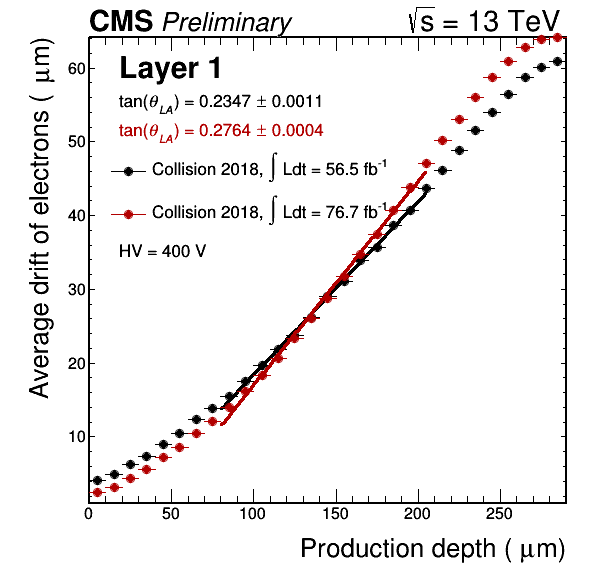}
    \caption{Lorentz angle measurements at integrated luminosities of 56.5 fb$^{-1}$ and 76.7 fb$^{-1}$.}
    \label{fig:Fig2}
\end{figure}

This parameter is sensitive to radiation effects. Radiation damage of the detector leads to higher $\theta_{LA}$ values. The Lorentz angle is measured by fitting the average drift distance of electrons as a function of production depth, as shown in Fig.\,\ref{fig:Fig2}. The plot shows two states of the sensor, one at the integrated luminosity of 56.5 fb$^{-1}$ and the other at 76.7 fb$^{-1}$. The $\theta_{LA}$ values throughout the history of the detector are stored in databases.

For a non-irradiated, fully depleted detector, the pixel charge profile is expected to be flat as the detector is fully efficient and all charge is collected, while for an irradiated detector losses are expected due to the trapping of carriers.

To cope with the irradiation one can either
\begin{itemize}
    \item anneal the detector, 
    \item increase the bias voltage, or 
    \item use a special reconstruction and simulate radiation damage in Monte Carlo simulations. 
\end{itemize}

\begin{figure}[h!]
    \centering
    \includegraphics[width=.6\textwidth]{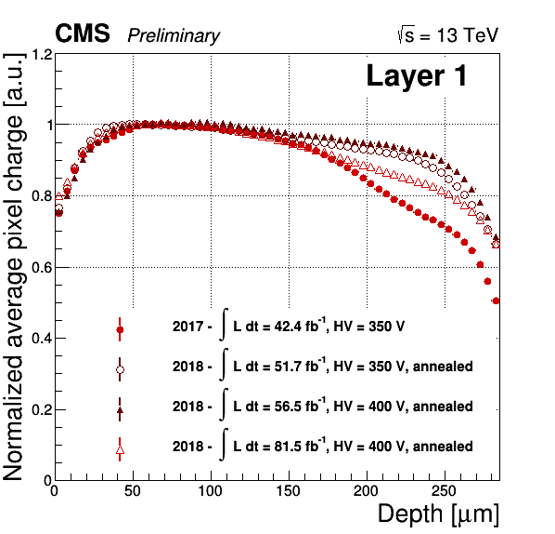}
    \caption{Cluster charge profiles.}
    \label{fig:Fig3}
\end{figure}

During the 2017 Extended Year Technical Stop, the barrel pixel detector was annealed at a temperature > 10 $^\circ$C for 53 days. The beneficial effect of the annealing during this period is clearly visible in the flattening of the pixel charge profile (Fig.\,\ref{fig:Fig3}). At the beginning of 2018 data taking, the charge collection was additionally increased in Layer 1 by raising the bias voltage from 350 V to 400 V (Fig.\,\ref{fig:Fig3}).

The special reconstruction method mentioned above simulates irradiated sensors in PixelAV\,\cite{Morris} corresponding to different charge profiles and stores these as 1D projections, called templates. This technique leads to superior resolution for irradiated sensors. The 2D projection is used from 2018 to reweight pixel charges in simulation, thus simulating the radiation damage of the silicon bulk.

\section{Resolution and Efficiency}
The main two parameters of a tracking detector are resolution and efficiency.

\begin{figure}[h!]
        \centering
    \includegraphics[width=.6\textwidth]{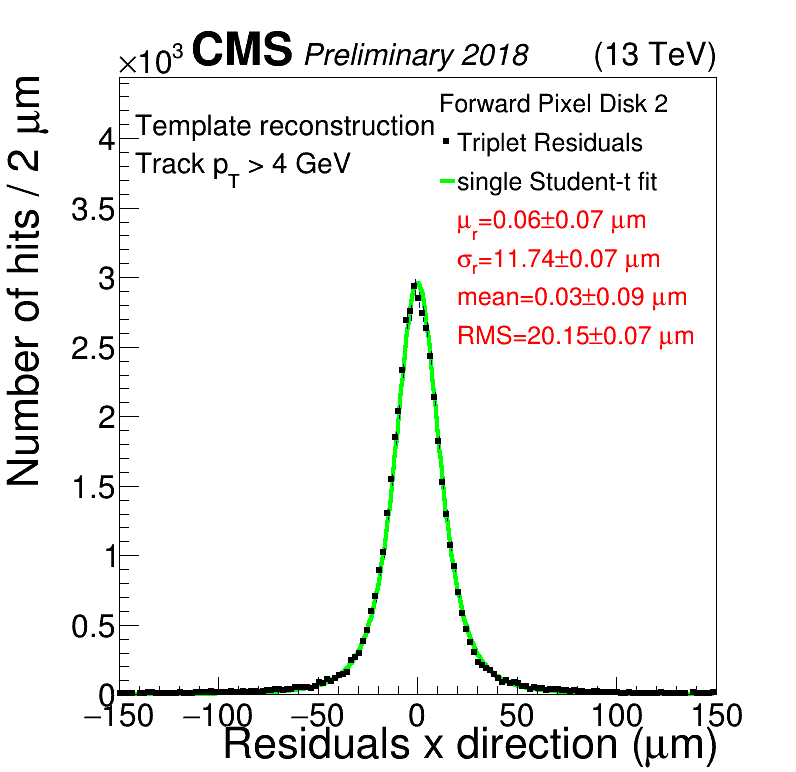}
    \caption{Residual distribution for disk 2.}
    \label{fig:Fig4}
\end{figure}

To measure the resolution, the so-called triplet method is used with the following steps (disk 2 is used as an example):
\begin{itemize}
    \item  Positions are reconstructed with the template    reconstruction algorithm.
    \item  Tracks with $p_{\mathrm{T}}$ > 4 GeV and hits in all three disks are selected and refitted using only hits in disks 1 and 3.
    \item  The trajectory is extrapolated to disk 2, and the residuals with respect to the actual hit are calculated.
    \item The residual distribution is fitted with the Student-t
    function.
    \item  Figure\,\ref{fig:Fig4} shows the residual distribution for disk 2, which has a width of 11.74 $\mu$m.
    \item The intrinsic resolution is extracted using simulations and it corresponds to about a factor of $\sqrt{2/3}$ of the width.
\end{itemize}

\newpage
Hit finding efficiency is defined as the fraction of all the projected trajectories where a matching cluster is found within a 1 mm radius (nearest cluster) around the expected trajectory position.

\begin{figure}[h!]
    \centering
    \includegraphics[width=.7\textwidth]{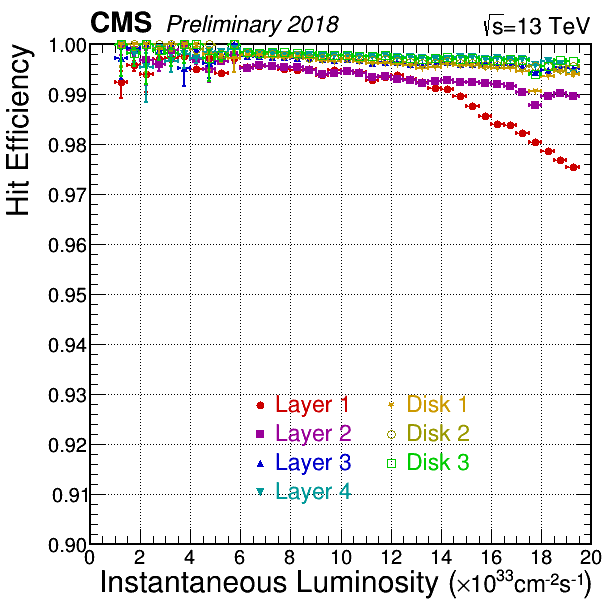}
    \caption{Hit efficiency as a function of integrated luminosity.}
    \label{fig:Fig5}
\end{figure}

The main reason for loosing hit efficiency is the saturation of the readout buffer in the chip. This is mostly relevant for high instantaneous luminosity and for the innermost layer. The Phase-1 pixel detector was designed to cope with this effect. For an instantaneous luminosity of 2$\cdot10^{34}$\,cm$^{-2}$s$^{-1}$ the efficiency for layer 1 is 97.5\,\%, while the efficiency for the other layers and the disks is above  99\,\%, as shown in Fig.\,\ref{fig:Fig5}.

\begin{figure}[h!]
    \centering
    \includegraphics[width=.6\textwidth]{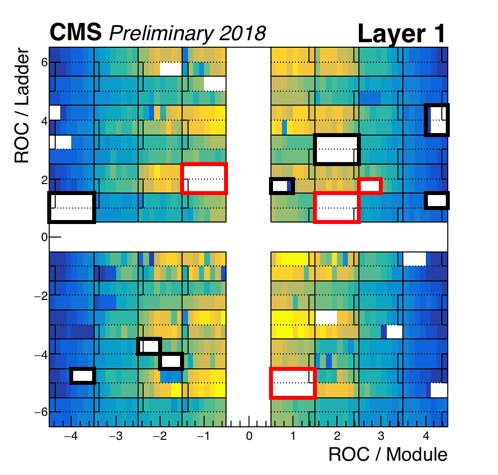}
    \caption{Occupancy map of layer 1.}
    \label{fig:Fig6}
\end{figure}

The efficiency definition above does not contain the temporarily or permanently bad detector elements. Figure\,\ref{fig:Fig6} shows the occupancy map of layer 1, where the white areas are the non-functioning parts. The modules with the black line correspond to bad components which were bad in 2017 as well, while the modules with the red line correspond to the new bad components. White areas without any line around them correspond to temporarily bad detector elements (SEUs), which are fixed at the reconfiguration of the chips.

From 2018 onwards a Prompt Calibration Loop \cite{PCL} based technique is used to determine bad components based on a dynamic occupancy threshold for each luminosity section. This information is stored in databases and propagated to the track reconstruction.

From 2019 onwards a dynamical bad component loss is simulated as well, including the simulation of the SEUs.

\newpage
\section{Impact parameter}
\begin{figure}[h!]
    \centering
    \includegraphics[width=\textwidth]{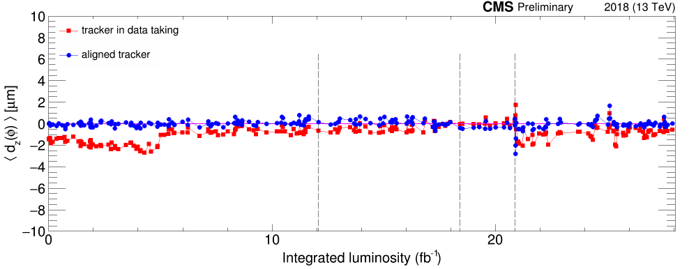}
    \caption{The average longitudinal impact parameter as a function of integrated luminosity.}
    \label{fig:Fig7}
\end{figure}

The average longitudinal impact parameter is shown in Fig.\,\ref{fig:Fig7}, as a function of integrated luminosity.

The vertical black lines indicate changes in the calibration of the local hit reconstruction. The red points show the results with the alignment constants used during data taking, the blue points show the results with the alignment constants as obtained in the offline alignment procedure.

Aligning the tracker improves the mean of this distribution.  Outliers in the trend are understood as degraded tracking performance caused by suboptimal pixel local reconstruction calibration input. Recalibrations for the whole Run 2 data are in progress (called Legacy reprocessing), which will further improve these results.

\section{Conclusions}

The pixel detector captures events every 25 nanoseconds using 124 million pixels. The resolution of the detector is a tenth of the size of an individual pixel and the efficiency of it is above 99\% expect for layer 1 at high luminosities.

To achieve this several calibrations are needed. Although in Run 2 the performance was already close to design specifications, it will further improve using the Legacy reprocessing.


\end{document}